\documentclass[prb,twocolumn,showpacs,preprintnumbers,amsmath,amssymb,superscriptaddress]{revtex4}
\usepackage{epsf}
\usepackage{graphicx}
\usepackage{bm}
\usepackage{amsmath}
\usepackage{color}
\usepackage{notes2bib}
\DeclareMathOperator{\sign}{sign}
\bibnotesetup{
note-name = , use-sort-key = false }
\begin{document}

\title{Indirect exchange interaction between magnetic adatoms in graphene}

\author{I.V. Krainov}
\email{igor.kraynov@mail.ru} \affiliation{Ioffe Institute, 194021 St. Petersburg,
Russia} \affiliation{Lappeenranta University of Technology, P.O. Box 20,
FI-53851, Lappeenranta, Finland}
\author{I.V. Rozhansky}
\affiliation{Ioffe Institute, 194021 St. Petersburg,
Russia} \affiliation{Peter the Great Saint-Petersburg Polytechnic University, 195251 St. Petersburg, Russia} \affiliation{Lappeenranta
University of Technology, P.O. Box 20,
FI-53851, Lappeenranta, Finland}
\author{N.S. Averkiev}
\affiliation{Ioffe Institute, 194021 St. Petersburg,
Russia}
\author{E. L\"ahderanta}
\affiliation{Lappeenranta University of Technology, P.O. Box 20,
FI-53851, Lappeenranta, Finland}


\date{\today}

\begin{abstract}
We present a theoretical study of indirect exchange interaction between magnetic adatoms
in graphene. The coupling between the adatoms to a graphene sheet is described in the framework
of tunneling Hamiltonian. We account for the possibility of this coupling being of resonant
character if a bound state of the adatom effectively interacts with the continuum of 2D delocalized states
in graphene. In this case the indirect exchange between the adatoms mediated by the 2D carriers appears to
be substantially enhanced compared to the results known from Ruderman-Kittel-Kasuya-Yosida (RKKY) theory.
Moreover, unlike the results of RKKY calculations in the case of resonant exchange the magnetic coupling
between the adatoms sitting over different graphene sublattices do not cancel each other.
Thus, for a random distribution of the magnetic adatoms over graphene surface a non-zero magnetic interaction is
expected. We also suggest an idea of controlling the magnetism by driving the tunnel coupling in and out of resonance
by a gate voltage.
\end{abstract}

\maketitle
\section{Introduction}
Magnetism in graphene has been attracting quite a lot of interest, probably no less
than any other physical property of this novel material.
In particular, quite a few experiments are focused on introducing magnetic properties by doping graphene with magnetic atoms
\cite{StmExp,PRBexp1,PRLexp2,PRBexp3}.
The magnetic adatoms deposited onto graphene surface in a moderate sheet density
can couple to the graphene and participate in the indirect exchange interaction
mediated by the free carriers available in graphene thus providing an analog of
a dilute magnetic semiconductor.
The indirect exchange is usually treated with RKKY theory which considers the spin-spin interaction
between impurities mediated by delocalized carriers.
As we have shown in our previous works
RKKY approach needs to be modified in the case when the impurity is coupled to the
free electron gas by means of resonant tunnelling \citep{OURPRB}.
Basically, the spatial separation between the magnetic center and free carriers
leads to decrease of the indirect exchange due to weak wavefunctions overlap.
However, if the magnetic center posess a bound state with an energy level matching the delocalized carriers energy range, the indirect exchange is dramatically increased due to resonant tunneling effect \citep{OURPss,OURJMMM}.
This phenomenon
was observed
in GaAs based heterostructures having an In$_x$Ga$_{1-x}$As quantum well (QW)
and Mn $\delta$-layer in the vicinity of QW \citep{Aron08,Aron10,Aron11,OURAPL2015}.

In this paper we analyse pair indirect exchange interaction
between magnetic adatoms via delocalized carriers in graphene.
For the non-resonant case when the bound state energy does not match the 2D continuum energy range
this approach resides to conventional RKKY theory in graphene with
some minor model-specific adjustments. The RKKY interaction in graphene has been intensively studied theoretically
\cite{Cryst,RKKY2007a,RKKY2007b,RKKY2010,Krash,RKKY2011a,RKKY2011b}.
It was shown that
the indirect exchange crucially depends on the magnetic centers position with regard to the graphene sublattices A and B.
If the magnetic centers are located at the vertices of the same sublattice the interaction is ferromagnetic while for the opposite
sublattices or plaquette configuration it is antiferromagnetic.
For the undoped graphene the interaction energy decreases as\cite{RKKY2007a,RKKY2007b,RKKY2010,RKKY2011a,Cryst} $1/R^3$ with distance $R$ between the impurities,
while for doped graphene the dependence is\cite{RKKY2007b,RKKY2011b} $1/R^2$.
For the case of magnetic adatoms being weakly coupled to the graphene the RKKY interaction is simply damped
by a factor of $\exp\left(-4d/d_0\right)$ with $d$ being the distance between an adatom and the graphene and $d_0$ characterizing decay
of the graphene wavefunction in the direction normal to its surface.
On the contrary, if an impurity has a resonant bound state the situation changes dramatically.
In the resonant case the energy of the indirect exchange interaction
is strongly enhanced and the interaction type (ferromagnetic or antiferromagnetic) can be the same both for AA and AB adatom configurations as
will be shown below.
\section{Theory}
Let us consider two magnetic adatoms located on a graphene sheet as shown in
Fig.1. Magnetic adatoms have spins labelled as $I_1,I_2$ and the bound states
with energy levels $\epsilon_1,\epsilon_2$.
We assume a tunnel coupling
which leads to a hybridization of the adatoms bound states descrete energy levels and
graphene continous spectrum. Such hybridization is described by the well-known Fano-Anderson model\cite{Fano}.
For a non-resonant case when the the discrete level energy lies outside of the occupied continuum spectrum
the perturbation theory of indirect exchange (RKKY) can be applied using the hybridized wavefunctions as the ground state.
However, if the discrete level energy lies within the 2D continuum the interaction becomes resonant, i.e. the
substantial part of the hybridized wavefunction resides at the magnetic centers.
At that, even a small perturbation of the bound state reflects in a strong change of the hybridized wavefunction.
This effect does not allow one to apply the RKKY theory which technically leads to a divergence of the energy correction
due to resonant denominators in the second order perturbation expressions.
An alternative approach, which avoids using the perturbation theory was suggested in our previous works
\cite{OURPRB,OURPss,OURJMMM}.
The approach still assumes weak exchange coupling regime so the magnetic adatoms spins are
treated as classical magnetic moments.
\begin{figure}[th]
  \centering
  \includegraphics[width=\linewidth]{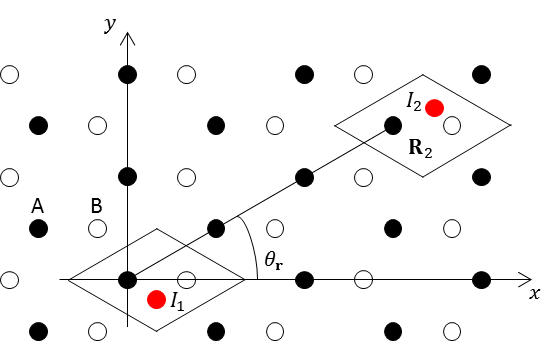}
  \caption{(Color online) The illustration of indirect exchange interaction of
  adatoms located on graphene. Two adatoms have spins $I_1, I_2$ and may be
  coupled to the bought graphene sublattice.}
  \label{scheme}
\end{figure}

We consider the following Hamiltonian:
\begin{equation}
\label{Htot}
\hat{H} = \hat{H}_0  + \hat{H}_T  + \hat{H}_J,
\end{equation}
where $\hat{H}_0$ describes the non-interacting adatoms and graphene,
$\hat{H}_T$  is the
tunneling term, which is spin independent, $\hat{H}_J$ describes the exchange interaction between the adatoms core spins and
the graphene electrons which have tunneled to the adatom.
We assume there is no direct exchange interaction between the adatoms.
For the weak coupling the adatoms spins act as parameters defining the potential
energy profile, thus the Hamiltonian (\ref{Htot}) becomes different for parallel and antiparallel spin configuration.
The difference in the system energy for parallel and antiparallel spin configurations of the magnetic adatoms can be interpreted as the indirect exchange interaction energy\cite{RKKY2010,Cryst,Krash}:
\begin{equation}
\label{EexchDef}
E_{ex} = E_{\uparrow \uparrow} - E_{\uparrow \downarrow}.
\end{equation}
We assume the exchange hamiltonian $\hat H_J$ as a contact interaction at the adatoms sites:
\begin{equation}
\hat{H}_J = A \delta({\bf r} - {\bf R}_1) \hat{\bf S} {\bf I}_1 + A \delta({\bf r} - {\bf R}_2) \hat{\bf S} {\bf I}_2,
\end{equation}
where $\hat{\bf S}$ is the electron spin operator, ${\bf I}_{1,2}$ are the adatoms spins, $A$ is the exchange constant, ${\bf R}_{1,2}$
denote positions of the two adatoms.
We express the whole Hamiltonian (\ref{Htot}) in the second quantization as follows:
\begin{align}
\label{HtotTerm}
 \hat{H}_{0} =& \varepsilon_1 \hat{f}_1^+ \hat{f}_1 + \varepsilon_2 \hat{f}_2^+
 \hat{f}_2 + v_F \sum_{\bf{p}} \bigl( |{\bf{p}}| \hat{c}_{{\bf{p}} K}^+
 \hat{c}_{{\bf{p}} K} - \nonumber\\
&- |{\bf{p}}| \hat{d}_{{\bf{p}} K}^+ \hat{d}_{{\bf{p}} K} + |{\bf{p}}|
\hat{c}_{{\bf{p}} K'}^+ \hat{c}_{{\bf{p}} K'} - |{\bf{p}}| \hat{d}_{{\bf{p}}
K'}^+ \hat{d}_{{\bf{p}} K'} \bigl), \nonumber \\
 \hat{H}_T =& t_{1A} \hat{a}_{R_1}^+ \hat{f}_1 + t_{1B} \hat{b}_{R_1}^+ \hat{f}_1
 + t_{2A} \hat{a}_{R_2}^+ \hat{f}_2 + t_{2B} \hat{b}_{R_2}^+ \hat{f}_2 + h.c.,
 \nonumber \\
 \hat{H}_J =& \lambda_1 \hat{f}_1^+ \hat{f}_1 + \lambda_2 \hat{f}_2^+ \hat{f}_2.
\end{align}
Here $\varepsilon_{1,2}$ are localized states energies, $\hat{c}_{{\bf{p}} K}^+,
\hat{d}_{{\bf{p}} K}^+, \hat{c}_{{\bf{p}} K'}^+, \hat{d}_{{\bf{p}} K'}^+$, $\hat{c}_{{\bf{p}} K},
\hat{d}_{{\bf{p}} K}, \hat{c}_{{\bf{p}} K'}, \hat{d}_{{\bf{p}} K'}$ are
the creation and annihilation operators of electrons (c) and holes (d) with discrete momentum ${\bf{p}}$
at $K$ and $K'$, points respectively. $v_F$ is the Fermi velocity in graphene,
$\lambda_{1,2} = A S I_{1,2} |\varphi(0)|^2$, where $S$, $I_{1,2}$ are the electron and the adatoms spin projections respectively,
$\varphi(0)$ is the bound state wave function at the magnetic adatom site.

To account for the coupling of each adatom to the graphene we introduced two complex parameters $t_{iA}, t_{iB},\, i=1,2$,
describing the tunnel coupling of $i$th adatom to the $A$ and $B$ graphene sublattice, respectively.
With that the tunneling part $\hat H_T$ in (\ref{HtotTerm}) is written in real space representation,
$\hat{a}_{R_i}^+, \hat{b}_{R_i}^+$ being the creation operators for real-space states at $A$ and $B$ sublattice.
Along with the part of the Hamiltonian $\hat{H}_0$ (\ref{HtotTerm}) describing graphene
the tunneling part must be rewtitten in the momentum representation.
Using tight-binding graphene Hamiltonian \cite{NovoselovRMP} operators in real-space
are expressed through those in reciprocal k-space as follows:
\begin{align}
\label{operators}
\hat{a}_{R_n} = & e^{-i{\bf{K}}{\bf{R}}_n} \sum_{\bf{p}}
\frac{e^{-i{\bf{p}}{\bf{R}}_n}}{\sqrt{N}} e^{-i\theta_{\bf{p}}}
\frac{\hat{c}_{\bf{p}}^{K} + \hat{d}_{\bf{p}}^{K}}{\sqrt{2}} + \nonumber \\
& + e^{-i{\bf{K}}'{\bf{R}}_n} \sum_{\bf{p}}
\frac{e^{-i{\bf{p}}{\bf{R}}_n}}{\sqrt{N}} e^{i\theta_{\bf{p}}}
\frac{\hat{c}_{\bf{p}}^{K'} + \hat{d}_{\bf{p}}^{K'}}{\sqrt{2}}, \nonumber \\
\hat{b}_{R_n} = & - e^{-i{\bf{K}}{\bf{R}}_n} \sum_{\bf{p}}
\frac{e^{-i{\bf{p}}{\bf{R}}_n}}{\sqrt{N}} \frac{\hat{c}_{\bf{p}}^{K} -
\hat{d}_{\bf{p}}^{K}}{\sqrt{2}} - \nonumber \\
& - e^{-i{\bf{K}}'{\bf{R}}_n} \sum_{\bf{p}}
\frac{e^{-i{\bf{p}}{\bf{R}}_n}}{\sqrt{N}} \frac{\hat{c}_{\bf{p}}^{K'} -
\hat{d}_{\bf{p}}^{K'}}{\sqrt{2}},
\end{align}
where ${\bf{R}}_n$ is a position of a carbon atom in the sublattice $A$,
$\theta_{\bf{p}}$ is a polar angle of the momentum vector ${\bf{p}}$ (here the coordinate
system is the same as in \cite{NovoselovRMP}). We assume that the first adatom is located at the
the origin ${\bf{R}}_1 = (0,0)$ and the second adatom is located at ${\bf{R}}_2
= (x,y) = (r, \theta_{\bf{r}})$ in Cartesian and polar coordinate system,
respectively.


We further proceed with diagonalization of  (\ref{Htot}) on a discrete basis in the same manner as described in \cite{OURJMMM}.
It appears to be more convenient to use a cylindrical discrete basis as the boundary conditions for the graphene in these coordiantes are simpler to define.
Let us put the system in a finite cylindrical box with radius $L$. Then one could specify a basis of the pure graphene eigenfunctions
using any type of boundary conditions. The graphene eigenfunctions in
cylindrical basis are:
\begin{align}
\label{psiK1c}
\psi^K_{p,n} (r,\theta_{\bf{R}}) = & \frac{\sqrt{p}}{2 \sqrt{L}} \begin{pmatrix}
J_n(pr) e^{i n \theta_{\bf{R}}} \\
\pm i J_{n+1}(pr) e^{i (n+1) \theta_{\bf{R}}}
\end{pmatrix} \nonumber\\
\psi^{K'}_{p,n} (r,\theta_{\bf{R}}) = & \frac{\sqrt{p}}{2 \sqrt{L}}
\begin{pmatrix}
J_n(pr) e^{i n \theta_{\bf{R}}} \\
\mp i J_{n-1}(pr) e^{i (n-1) \theta_{\bf{R}}}
\end{pmatrix},
\end{align}
where up and down sign correspond to the electrons and holes, respectively.
We take the boundary conditions for $K$ and $K'$ valleys in a general form
as discussed in \cite{Volkov}:
\begin{align}
\hat{\Gamma}_K = & \begin{pmatrix}
1 && i \beta e^{-i \theta_{\bf{R}}} \\
0 && 0
\end{pmatrix}, \hat{\Gamma}_{K'} = \begin{pmatrix}
1 && i \beta e^{i \theta_{\bf{R}}} \\
0 && 0
\end{pmatrix}, \nonumber \\
& \hat{\Gamma}_K \psi^K_{p,n} (r,\theta_{\bf{R}}) \Bigl|_{r=L} = 0, \nonumber \\
\label{bcK2}
& \hat{\Gamma}_{K'} \psi^{K'}_{p,n} (r,\theta_{\bf{R}}) \Bigl|_{r=L} = 0,
\end{align}
here  parameter $\beta$ describes the boundary. Using
(\ref{bcK2}) we get discrete energy levels for $K$ and $K'$ valleys:
\begin{equation}
|E_{m,n}^0| =
\hbar v_F \left( \frac{\pi m }{ L} + \frac{\pi (2 n + 1) }{ 4 L} \pm \arctan \frac{\beta}{ L} \right),
\end{equation} where
$"+"$ corresponds to the holes and $"-"$ corresponds to the electrons,
$m,n$ are integer values, $m$ is the quantized momentum amplitude $n$ is the cylindrical harmonic number.
Now let us consider the total Hamiltonian (\ref{Htot}),
we write its eigenfunctions $|\Psi\rangle$ as an expansion:
\begin{align}
\label{psiNew}
|\Psi\rangle = & \bigl( a_1 \hat{f}_{1}^+ + a_2 \hat{f}_{2}^+ + b_{p n}^{c K}
\hat{c}_{p n K}^+ + b_{p n}^{d K} \hat{d}_{p n K}^+ + \nonumber \\
& + b_{p n}^{c K'} \hat{c}_{p n K'}^+ + b_{p n}^{d K'} \hat{d}_{p n K'}^+ \bigl)
|0\rangle,
\end{align}
where the coefficients $a_i,b_\alpha^\beta$ are to be determined. The indexes $p, n$ characterize the cylindrical basis states (\ref{psiK1c}).
The tunnelling matrix elements in the cylindrical basis (\ref{psiK1c}) could
be obtained from those in the plane waves basis (\ref{operators}):
\begin{equation}
\label{tunMatr}
t_{1 p n}^{c K} = i^{-n} \sqrt{ \frac{\pi L p}{2}} \int_0^{2 \pi} \frac{d
\theta_{{\bf{p}}}}{2 \pi} e^{-i n \theta_{{\bf{p}}}} \langle 0 |\hat{f}_1
\hat{H}_T \hat{c}^+_{{\bf{p}} K} | 0 \rangle.
\end{equation}
Finally, in the basis
\[(\hat{f}_{1}^+ |0\rangle, \hat{f}_{2}^+ |0\rangle, \{ \hat{c}_{p n
K}^+ |0\rangle \}, \{ \hat{d}_{p n K}^+ |0\rangle \}, \{ \hat{c}_{p n K'}^+
|0\rangle \}, \\ \{ \hat{d}_{p n K'}^+ |0\rangle\} )\] the Hamiltonian (\ref{Htot})
reads:
\begin{equation}
\label{Hmatrix}
\hat{H} =
\begin{pmatrix}
\varepsilon_1 + \lambda_1 & 0 & t_{1 p n}^{c K} & t_{1 p n}^{d K} & t_{1 p n}^{c
K'} & t_{1 p n}^{d K'} \\
0 & \varepsilon_2 + \lambda_2 & t_{2 p n}^{c K} & t_{2 p n}^{d K} & t_{2 p n}^{c
K'} & t_{2 p n}^{d K'} \\
t_{1 p n}^{c K *} & t_{2 p n}^{c K *} & E_{p n}^c & 0 & 0 & 0 \\
t_{1 p n}^{d K *} & t_{2 p n}^{d K *} & 0 & E_{p n}^d & 0 & 0 \\
t_{1 p n}^{c K' *} & t_{2 p n}^{c K' *} & 0 & 0 & E_{p n}^c & 0 \\
t_{1 p n}^{d K' *} & t_{2 p n}^{d K' *} & 0 & 0 & 0 & E_{p n}^d \\
\end{pmatrix}.
\end{equation}
The eigenvalues of $\hat{H}$ (\ref{Hmatrix}) can be found analytically, the details are presented in Appendix A.
The indirect exchange energy is the difference between parallel and antiparallel adatoms spin configurations, in order to find the total energy difference
one should sum up the difference over all occupied states:
\begin{align}
\label{EexcCalc}
E_{ex} =\sum_{m,i} \sum_{s=\pm1/2}\bigl[ E_{m,i}(\uparrow\uparrow,s)
 - E_{m,i}(\downarrow\uparrow,s)\bigl],
\end{align}
where $m=0,1,2,...$ and $i=1,2$ characterise the modified energy levels (eigenvalues of (\ref{Hmatrix})) and the outer sum is over all occupied energy levels (see Appendix A for details).
Also for each of the adatoms configurations ($\uparrow\uparrow$ denotes parallel spin configuration, $\uparrow\downarrow$ is for antiparallel spin configuration) the summation is done over the electron spin projection $s$.
The adatoms spin configurations and electron spin prohection enter (\ref{Hmatrix}) only through the parameters $\lambda_{1,2}$, which are of the same
magnitude $|\lambda_1| = |\lambda_2|\equiv\lambda$ but can have different sign, for example for $\uparrow\uparrow, s=-1/2$ these parameters are
$\lambda_1=\lambda_2=-\lambda$.
In the following we keep the leading order in the tunnelling parameters. As shown in Appendix B the indirect exchange energy contains
a fast oscillating factors with a characteristic spatial period $2\pi/K$, that is the length
of the carbon-carbon bond, the same applies to a conventional RKKY interaction in graphene \cite{RKKY2010,RKKY2011a}.
Here we average over these short-wavelength oscillations and also average on the polar angle of
describing the adatoms position (for details see Appendix B).
Finally, we get for the resonant indirect
exchange interaction in graphene:
\begin{align}
\label{EexcAns}
&E_{ex}  = \int\limits_{-\infty}^{E_F} \frac{d E}{\pi} \arctan
\left\{ \frac{\lambda^2 E^2
f(E,r)}{\left[(\varepsilon_1-E)^2-\lambda^2\right]\left[(\varepsilon_2-E)^2-\lambda^2\right]}
\right\} \cdot \nonumber \\
&\ \ \ \cdot \sign(E), \\
\label{fDef}
&f(E,r) = \tau^4_{AA} J_0\left(\frac{|E| r}{\hbar v_F}\right)N_0\left(\frac{|E|
r}{\hbar v_F}\right)+ \nonumber \\
&\ \ \ +\tau^4_{AB} J_1\left(\frac{|E| r}{\hbar v_F}\right)N_1\left(\frac{|E|
r}{\hbar v_F}\right), \\
&\tau^4_{AA} = \frac{16}{3 t^4}
(t_{1A}t_{2A}^*+t_{1B}t_{2B}^*)(t_{1A}^*t_{2A}+t_{1B}^*t_{2B}), \nonumber \\
& \tau^4_{AB} = \frac{16}{3 t^4} (|t_{1A}|^2|t_{2B}|^2+|t_{1B}|^2|t_{2A}|^2),
\nonumber
\end{align}
where $t\approx 2.8~$eV - is the graphene's nearest-neighbor hopping energy
\cite{NovoselovRMP}, $J_i, N_i$ - are Bessel and Neumann functions of
$i$th order, respectively.
\section{Discussion}
Let us assume for simplicity that the adatoms bound states
have the same localization energy, $\varepsilon_1 =
\varepsilon_2\equiv \varepsilon_0 $.
For the resonant case, when the bound state energy lies within the 2D continuum spectra, $\varepsilon_0 \in (-\infty,E_F)$,
from (\ref{EexcAns}) the exchange energy can be roughly estimated as:
\begin{equation}
\label{Eres}
E_{res} \sim \frac{T}{t} \sqrt{\lambda \varepsilon_0},\,\,\,\,  T=\left(\tau^4_{AA}+\tau^4_{AB}\right)^{1/4}
\end{equation}
To consider the
non-resonant case let us assume the localized state energy lying above the Fermi level,
$\varepsilon_0 \gg E_F$.
In this case the arctangent argument (\ref{EexcAns}) has no poles and
$\arctan$ is to be replaced by its argument (due to the smallness of the tunneling parameter).
Further calculations are similar to those discussed
in  \cite{RKKY2007a}.
We get:
\begin{align}
\label{Enr}
E_{nr} \approx& \frac{\lambda^2}{\pi \varepsilon_0^4} \int
\limits_{-\infty}^0 dE\ E^2 f(E,r) = \nonumber \\
& = \frac{27}{128 \pi} \frac{\lambda^2}{t} \frac{1}{(r/a)^3} \left(
\frac{t}{\varepsilon_0} \right)^4 \left[ -\tau_{AA}^4 + 3 \tau_{AB}^4 \right].
\end{align}
\begin{figure}[th]
  \centering
  \includegraphics[width=\linewidth]{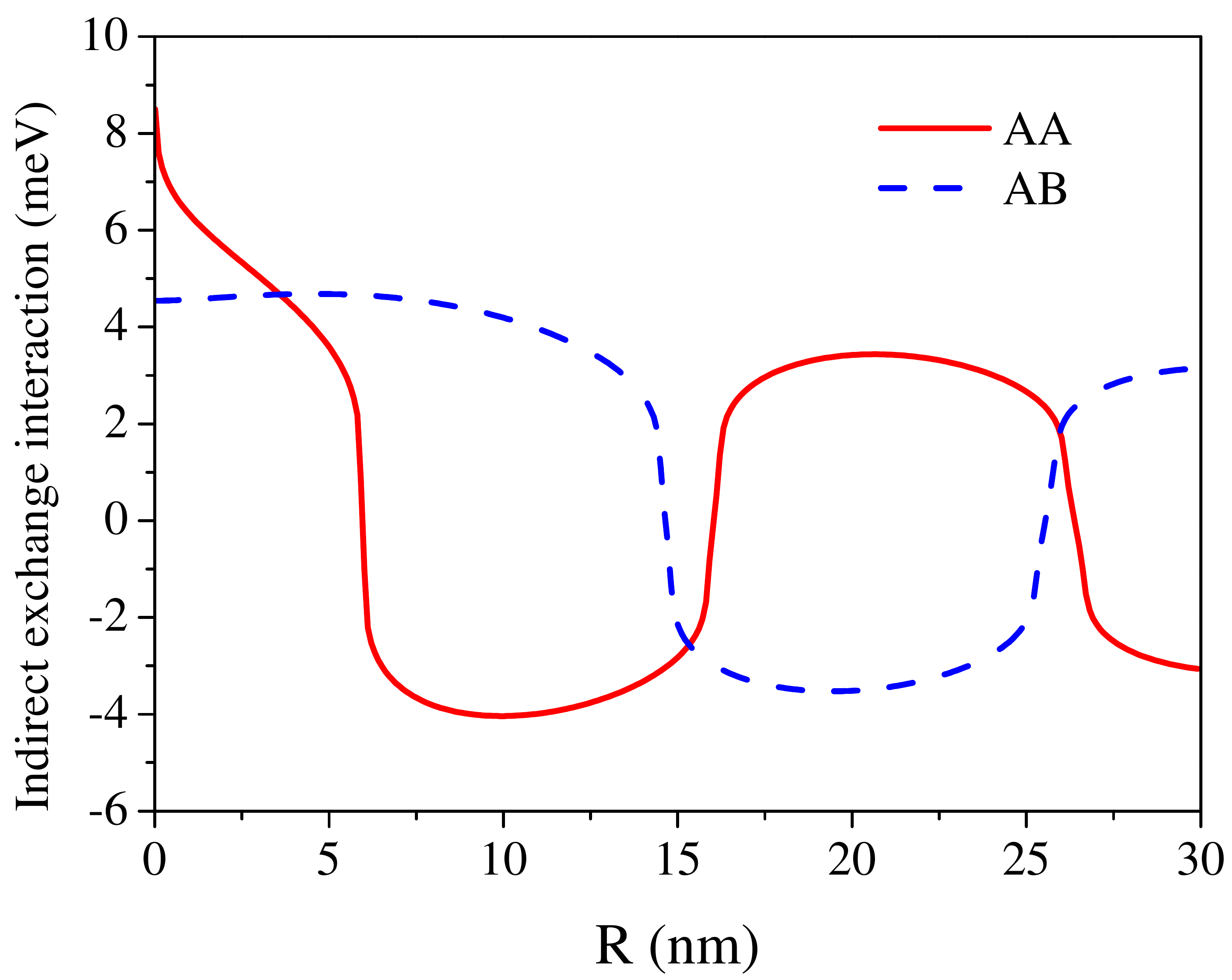}
  \caption{(Color online) Indirect exchange interaction energy vs distance
  between adatoms. Localized state's energy is $\varepsilon_0 = -100~$meV, Fermi
  level is $E_F= 0~$meV. Red curve - adatoms located on same sublattice (AA),
  blue dash curve - adatoms located on different sublattice (AB).}
  \label{figEres}
\end{figure}
\begin{figure}[bh]
  \centering
  \includegraphics[width=\linewidth]{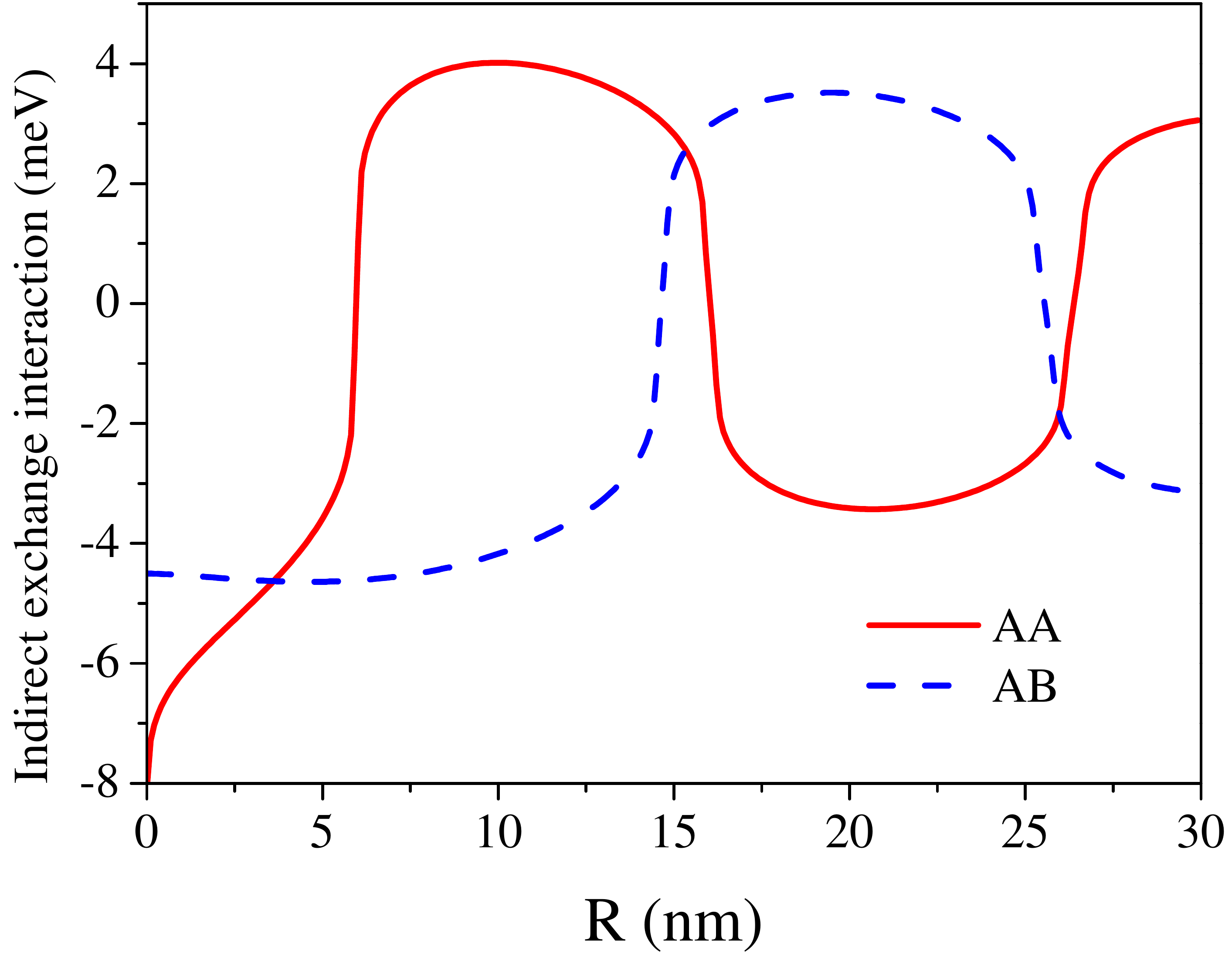}
  \caption{(Color online) Indirect exchange interaction energy vs distance
  between adatoms. Localized state's energy is $\varepsilon_0 = 100~$meV, Fermi
  level is $E_F= 110~$meV. Red curve - adatoms located on same sublattice (AA),
  blue dash curve - adatoms located on different sublattice (AB).}
  \label{figEmag}
\end{figure}
The expression (\ref{Enr}) is consistent with the previously known results of conventional RKKY theory in graphene
\citep{RKKY2007b,RKKY2010,RKKY2011a,RKKY2011b} with the characteristic dependence $E_{nr} \sim \lambda^2 T^4$.
While in metals the the RKKY-type indirect exchange interaction is usually ferromagnetic
at small distances between the magnetic impurities, in graphene it depends on the adatoms position. As
follows from (\ref{Enr}) if both adatoms are located at the same sublattice (AA or BB)
the interaction is ferromagnetic while for different subllatices configuration (AB, BA) it is antiferromagnetic.
This is in exact agreement with the RKKY theory in graphene \cite{RKKY2011a}.
However, for the resonant case a picture of the indirect exchange interaction appears to be radically different
from what is expected from the conventional RKKY theory.
In order to apply the resonant exchange theory to any real case one should be aware of
typical values of the parameters entering (\ref{EexcAns}). Those can be estimated as
 $\lambda \sim 1 \div 10~$meV \cite{OURPRB,OURAPL2015}, $T \sim 0.1~\div~1~$eV \cite{StmTheor,StmExp}.
The STM experiments \cite{StmExp} have shown that Co adatoms have localized
energy level at $\varepsilon_0 \sim -100~$meV and the Fermi level in graphene can be
effectively controlled by gate voltage in the range $E_F \sim -300~\div~300~$meV \cite{StmExp}.
Figs.\ref{figEres},\ref{figEmag} show the resonant indirect exchange interaction
as a function of the distance between the adatoms calculated according to the formula (\ref{EexcAns}) for different adatoms configurations (AA and AB).
For this calculation we took the following values:
$\lambda = 1~$meV.
For the AA configuration the tunneling parameters
describe the coupling of the adatoms to only one sublattice (A)
$t_{1A} = t_{2A} = 1~$eV,$\ t_{1B} = t_{2B} = 0$.
For the (AB) configuration we took
$t_{1A} = t_{2B} = 1~$eV,$\ t_{1B} = t_{2A} = 0$.
We analyse two different situations regarding the position of the bound state energy related
to the Dirac point: (i) $\varepsilon_0 = - 100~$meV, $E_F = 0~$meV, (ii) $\varepsilon_0 = 100~$meV, $E_F = 110~$meV.
In both cases the resonance condition is satisfied.
Case (i) is presented in Fig.\ref{figEres}. Note that in contrast to the conventional (non-resonant) RKKY theory the interaction between the adatoms is antiferromagnetic at small distances for both AA and AB configurations.
The reason is that unlike in the RKKY case
in the resonant case the tunnelling is enhanced for the graphene electrons having the same energy as the localized state
and for small distance both terms in (\ref{fDef}) appear to be of the same sign.
In the case (ii) the bound state energy level lies in the conduction band above the Dirac point.
The calculated indirect exchange energy is presented in  Fig.\ref{figEmag}.
In this case for both AA and AB configurations the interaction is ferromagnetic at a small distance.
Note, that because the contributions from AA and AB configurations do not compensate each other, the
plaquette configuration or random distributions of the magnetic adatoms would also result in a ferromagnetic interaction. Thus, the resonant exchange can be revealed in an experiment with a random distribution of the adatoms.
For both resonant cases (i) and (ii) discussed above the characteristic wavelength for the oscillations
is governed by the resonant energy: $\Lambda\sim a \cdot  t / \varepsilon_0$ while in conventional RKKY theory
it is related to the Fermi energy $\Lambda\sim v_F/\varepsilon_F$.

\begin{figure}[th]
  \centering
  \includegraphics[width=\linewidth]{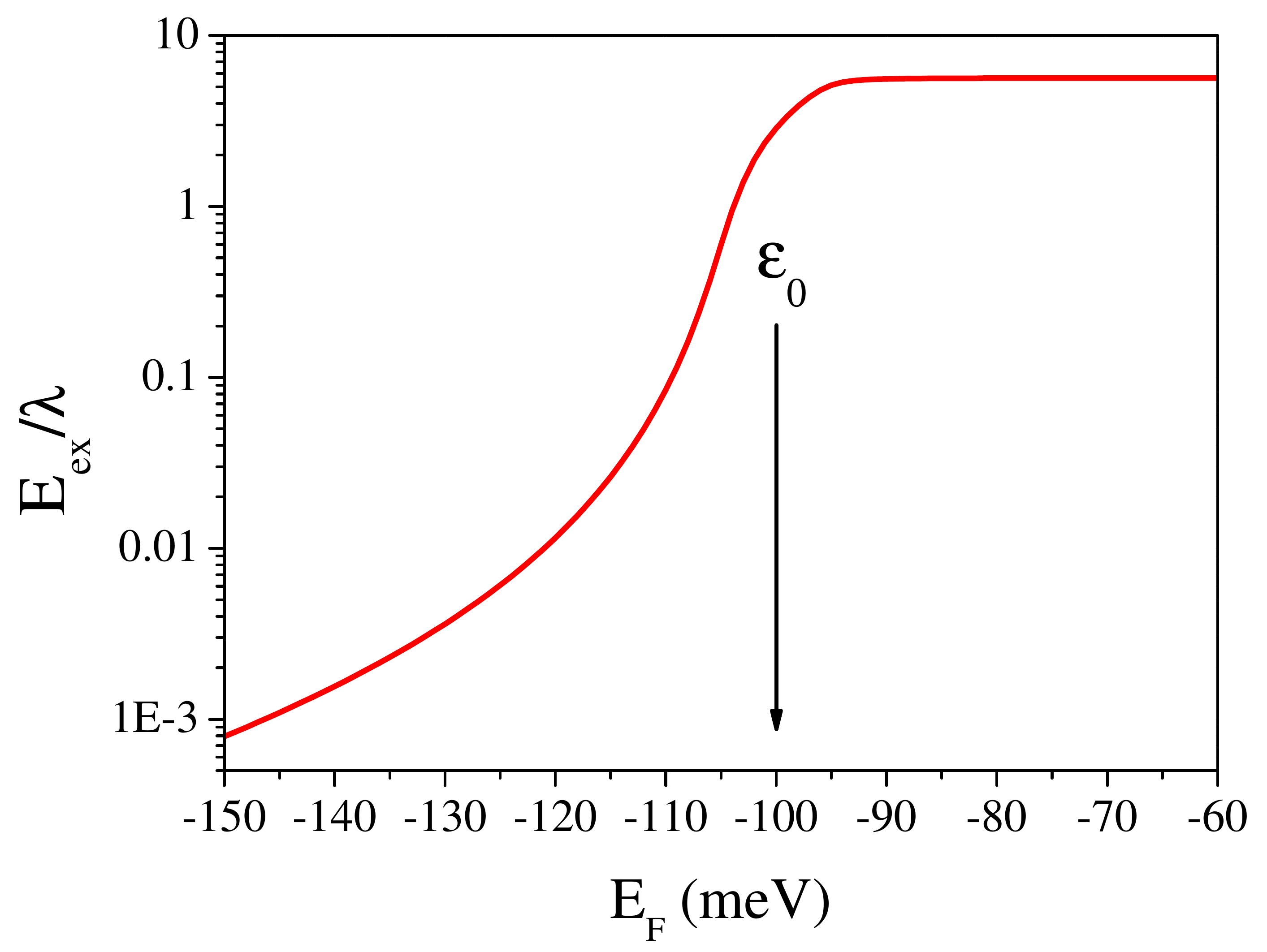}
  \caption{(Color online) Indirect exchange interaction energy as a
  function of the Fermi level position. Bound state energy is $\varepsilon_0 =
  -100~$meV, the distance between adatoms is $R = 2$ nm, AA configuration.}
  \label{figEexEf}
\end{figure}

An important feature of the resonant indirect exchange is
that its strength can be several orders of magnitude higher compared to the non-resonant case.
This effect has been described for semiconductor heterostructures \cite{OURPRB} and carbon nanotubes \cite{OURJMMM}.
The physical reason for such an enhancement is simple, the resonant tunneling makes the free carriers wavefunction effectively penetrates
onto the magnetic center where it participates in the direct exchange. The enhancement although not that large has been seen experimentally for hybrid semiconductor-Mn heterostructure \cite{OURAPL2015}.
Here we suggest a natural idea of modulating the magnetic properties of the system by adjusting the Fermi level in order
to drive the system in and out of the resonant condition.
To illustrate the effect we plotted the indirect exchange energy vs Fermi level position in Fig.\ref{figEexEf}.
The distance between the adatoms located on same sublattice (AA configuration) was kept $R = |{\bf R_1}-{\bf R_2}| =2~$nm and the bound state energy
was taken $\varepsilon_0=-100$ meV.
As seen from the Fig. \ref{figEexEf} the indirect exchange energy changes by several orders of magnitude while changing
the Fermi level position $E_F$ by $\sim100$ meV. Such a modulation of the Fermi level position can be rather easily achieved by applying the
appropriate gate voltage. Thus, along with the modulation of the interaction range\cite{Cryst} one could effectively control its strength using the gate voltage.
\section{Summary}
In conclusion, we have developed an indirect exchange interaction theory for magnetic
adatoms on graphene.
The presence of resonant localized states at the adatoms
strongly affects the picture of indirect interaction.
The indirect exchange is strongly enhanced whenever the adatom bound state energy
level lies within the energy range of occupied states in graphene.
Thus, the indirect interaction strength can be effectively controlled by adjusting
the Fermi level position in graphene with an external gate voltage.
An important property of the resonant exchange in graphene is that
unlike non-resonant RKKY theory in graphene the type of the interaction (ferromagnetic or
antiferromagnetic) is the same for the same-sublattiice (AA) and different sublattice (AB) configurations.
 The interaction is ferromagnetic on small distance if the bound state energy level lies in the conduction band and antiferromagnetic for the bound state energy energy level lying in the valence band.
Because the contributions from AA and AB configurations do not compensate each other in the resonant case, the random distributions of the magnetic adatoms results in a non-zero magnetic interaction. This finding
opens a good possibility for studying the phenomena experimentally.

\section{Acknowledgments}
This work has been carried out under the financial support of a
Grant from the Russian Science Foundation(Project no.14-12-
00255). Also, I.V.~Krainov thanks Dynasty Foundation.

\appendix

\section{Calculation of eigenvalues}
Putting aside some complications introduced by graphene, our
problem is the one
of interaction between two discrete levels and continuum, so-called Fano-Anderson problem \cite{Fano}.
The 'continuum' can be represented by a dense set of discrete levels, the
determinant of the appropriate eigenvalue problem can be reduced to
a 2x2 form:
\begin{align}
\label{detRed}
\begin{vmatrix}
\varepsilon_1 + A_1 + \lambda_1 - E && B \\
B^* && \varepsilon_2 + A_2 + \lambda_2 - E\\
\end{vmatrix},
\end{align}
where
\begin{align}
\label{AdefSum}
A_i = \sum_{m,n} &\Biggl( \frac{t_{ipn}^{cK}t_{ipn}^{cK*}}{E-E_{mn}^{0c}}+ \frac{t_{ipn}^{dK}t_{ipn}^{dK*}}{E-E_{mn}^{0d}}+ \frac{t_{ipn}^{cK'}t_{ipn}^{cK'*}}{E-E_{mn}^{0c}}+ \frac{t_{ipn}^{dK'}t_{ipn}^{dK'*}}{E-E_{mn}^{0d}} \Biggl),\\
\label{BdefSum}
B = \sum_{m,n} &\Biggl( \frac{t_{1pn}^{cK}t_{2pn}^{cK*}}{E-E_{mn}^{0c}}+ \frac{t_{1pn}^{dK}t_{2pn}^{dK*}}{E-E_{mn}^{0d}}+ \frac{t_{1pn}^{cK'}t_{2pn}^{cK'*}}{E-E_{mn}^{0c}}+ \frac{t_{1pn}^{dK'}t_{2pn}^{dK'*}}{E-E_{mn}^{0d}} \Biggl).
\end{align}
The summation in (\ref{AdefSum},\ref{BdefSum}) is over all basis states, $m,n = 0,1,2,...$.
 The quantized momentum $p$ in the tunnelling matrix element is discrete and corresponds to the index $m$ through the quantization condition (\ref{bcK2}).
Indexes $c,d$ correspond to the electrons and holes, respectively.
Equation (\ref{AdefSum}) describes a shift of the adatoms energy levels due to interaction with graphene carriers.
In the following calculation we change definition of the bound states energy levels as $\varepsilon_i + A_i
\rightarrow \varepsilon_i$. Note that the shift $A_i$ is of the order of $T^2$ and can be neglected in our case .
Equation (\ref{BdefSum}) describes interaction between adatoms via graphene electrons.
Expanding (\ref{detRed}) we get:
\begin{align}
\label{charPol}
&(\varepsilon_1 + \lambda_1 - E) (\varepsilon_2 + \lambda_2 - E) - B B^* = 0, \\
\label{Bdef}
B = & \frac{\Omega_{u.c.} E}{4 (\hbar v_F)^2} \Bigl\{ \left( e^{i {\bf{K}}
{\bf{R}}} + e^{i {\bf{K}}' {\bf{R}}} \right) \left(t_{1A} t_{2A}^* + t_{1B}
t_{2B}^* \right) \cdot \nonumber \\
& \cdot \left( N_0(kr) + J_0(kr) \cot(kL) \right) + i \Bigl[ \Bigl( e^{i
{\bf{K}} {\bf{r}} - i \theta_{{\bf{r}}}} + \nonumber \\
& + e^{i {\bf{K}}' {\bf{r}} + i \theta_{{\bf{r}}}} \Bigl) t_{1A} t_{2B}^* +
\left( e^{i {\bf{K}} {\bf{R}} + i \theta_{{\bf{r}}}} + e^{i {\bf{K}}' {\bf{R}} -
i \theta_{{\bf{r}}}} \right) \cdot \nonumber \\
& \cdot t_{1B} t_{2A}^* \Bigl] \left( N_1(kr) + J_1(kr) \cot(kL) \right)
\Bigl\},
\end{align}
where ${\bf{R}} = {\bf{R}}_2 - {\bf{R}}_1$, $k = E / \hbar v_F$,
$\Omega_{u.c.} = 3 \sqrt{3} a^2 / 2$ is the graphene unit cell area,
$a$ is the C-C bond length, $J_i, N_i$- are Bessel and Neumann functions of
$i$-th order, respectively.
The unit cell area emerges from (\ref{operators}) and (\ref{tunMatr}) as $L/N=\Omega_{u.c.}/2\pi L$.
The roots of the characteristic polynomial (\ref{charPol})
can be parameterized with indexes $E_{m,i}$, where $m$ describes the absolute value of momentum in unperturbed
state $E_m^0$, $i=1,2$ indicates two solutions of (\ref{charPol}) in accordance with the number of
adatoms considered (two).

\section{Exact equation for resonant indirect exchange interaction}
In this section we present an exact calculation of resonant indirect exchange
interaction following (\ref{charPol}). We obtained:
\begin{align}
\label{EexcAc}
&E_{ex} = \int\limits_{-\infty}^{E_F} \frac{d E}{\pi} \arctan \left\{
\frac{\lambda^2 E^2
g(E,r)}{\left[(\varepsilon_1-E)^2-\lambda^2\right]\left[(\varepsilon_2-E)^2-\lambda^2\right]}
\right\} \cdot \nonumber \\
&\ \ \ \cdot \sign(E) \\
\label{gDef}
&g(E,r) = \tau^4_{AA} \left( 1 + \cos\left({\bf{K}}-{\bf{K}}'\right){\bf{R}}
\right) J_0\left(\frac{|E| r}{\hbar v_F}\right)N_0\left(\frac{|E| r}{\hbar
v_F}\right)+ \nonumber \\
&\ \ \ +\bigl[ \tau^4_{1AB} \left( 1 + \cos
\left(\left({\bf{K}}-{\bf{K}}'\right){\bf{R}} - 2 \theta_{{\bf{r}}}\right)
\right) + \nonumber \\
& \ \ \ + \tau^4_{2AB} \left( 1 + \cos
\left(\left({\bf{K}}-{\bf{K}}'\right){\bf{R}} + 2 \theta_{{\bf{r}}}\right)
\right) + \nonumber \\
&\ \ \ + 2 \tau^2_{1AB}\tau^2_{2AB} \cos(\alpha_{1AB}-\alpha_{2AB}) (
\cos\left({\bf{K}}-{\bf{K}}'\right){\bf{R}} + \nonumber \\
&\ \ \ + \cos (2 \theta_{{\bf{r}}}) ) \bigl] J_1\left(\frac{|E| r}{\hbar
v_F}\right)N_1\left(\frac{|E| r}{\hbar v_F}\right)+ \nonumber \\
&\ \ \ + \bigl[ \tau^2_{AA} \tau^2_{1AB} \sin (\alpha_{AA}-\alpha_{1AB}) (\cos
\left(\left({\bf{K}}-{\bf{K}}'\right){\bf{R}} -  \theta_{{\bf{r}}}\right)) +
\nonumber \\
&\ \ \ + \tau^2_{AA} \tau^2_{2AB} \sin (\alpha_{AA}-\alpha_{2AB}) (\cos
\left(\left({\bf{K}}-{\bf{K}}'\right){\bf{R}} +  \theta_{{\bf{r}}}\right)) \bigl]
\cdot \nonumber \\
&\ \ \ \cdot \left(J_0\left(\frac{|E| r}{\hbar v_F}\right)N_1\left(\frac{|E|
r}{\hbar v_F}\right) + J_1\left(\frac{|E| r}{\hbar v_F}\right)N_0\left(\frac{|E|
r}{\hbar v_F}\right) \right), \nonumber \\
\end{align}
\begin{align}
&\tau^2_{AA} e^{i\alpha_{AA}} = \frac{4}{\sqrt{3} t^2}
(t_{1A}t_{2A}^*+t_{1B}t_{2B}^*), \nonumber \\
& \tau^2_{1AB} e^{i\alpha_{1AB}} = \frac{4}{\sqrt{3} t^2} t_{1A}t_{2B}^*,
\nonumber \\
& \tau^2_{2AB} e^{i\alpha_{2AB}} = \frac{4}{\sqrt{3} t^2} t_{1B}t_{2A}^*.
\nonumber
\end{align}
Let us consider the non-resonant limiting case. We assume  ($\varepsilon_0 \equiv \varepsilon_1 =
\varepsilon_2$), $\varepsilon_0 \gg E_F$, for undoped graphene we take $E_F=0$.
We analyse the two configurations: (i) adatoms are located on same sublattice $T_{AA} = t_{1A} = t_{2A}, \ t_{1B} = t_{2B} = 0$, (ii)
the adatoms are located on different graphene sublattice $T_{AB} = t_{1A} = t_{2B},\
t_{2A} = t_{1B} = 0$.
For the two sublattice configurations we have
\begin{align}
\label{EnrAA}
&E_{nr}^{AA} =  -\frac{9}{8 \pi} \frac{\lambda^2}{t}
\frac{1+\cos\left({\bf{K}}-{\bf{K}}'\right){\bf{R}}}{(R/a)^3} \left(
\frac{T_{AA}}{\varepsilon_0} \right)^4 \\
\label{EnrAB}
&E_{nr}^{AB} = \frac{27}{8 \pi} \frac{\lambda^2}{t} \frac{1+\cos
\left(\left({\bf{K}}-{\bf{K}}'\right){\bf{R}} - 2
\theta_{{\bf{r}}}\right)}{(R/a)^3} \left( \frac{T_{AB}}{\varepsilon_0} \right)^4
\end{align}
These results (\ref{EnrAA},\ref{EnrAB}) describe indirect exchange interaction in non-resonant limit
and are in agreement with those obtained using RKKY theory in graphene \citep{RKKY2007b,RKKY2010,RKKY2011a,RKKY2011b},
the difference in the pre-factor resembles the specifics of our model where the bound state level exist also in a non-resonant case.
Note that the absence of $\pi$
factor in exact equation for $g(E,r)$ (\ref{gDef}) is connected to the $\pi / 2$ angle
difference in our coordinate system from \cite{RKKY2010,RKKY2011a}. As one could
see from (\ref{EexcAc}) the indirect exchange oscillates with atomic
length period. We make an averaging over these short-range oscillations:
\begin{align}
\label{EavDef}
\langle E_{ex} \rangle &= \int\limits_0^{2\pi} \frac{d \gamma}{2\pi}
E_{ex}(\gamma), \\
\gamma &= \left({\bf{K}}-{\bf{K}}'\right){\bf{R}}. \nonumber
\end{align}
This integral cannot be calculated analytically, but we can neglect the
oscillation terms (i.e. $\cos(\gamma) \rightarrow 0$) in (\ref{gDef}). We checked
the error associated with disregarding these terms by a numerical calculation of (\ref{EavDef})
and found good agreement, the maximum error is less then 20\% \cite{RKKY2011b}.

\bibliography{RKKY}

\end{document}